# A Universal Model of Global Civil Unrest

Dan Braha

New England Complex Systems Institute, 238 Main St., Suite 319, Cambridge, MA 02142, USA

University of Massachusetts Dartmouth, Dartmouth, MA 02747, USA

(Dated: May 2012)

**Civil unrest is a powerful form of collective human dynamics, which has led to major transitions of societies in modern history. The study of collective human dynamics, including collective aggression, has been the focus of much discussion in the context of modeling and identification of universal patterns of behavior. In contrast, the possibility that civil unrest activities, across countries and over long time periods, are governed by universal mechanisms has not been explored. Here, we analyze records of civil unrest of 170 countries during the period 1919-2008. We demonstrate that the distributions of the number of unrest events per year are robustly reproduced by a nonlinear, spatially extended dynamical model, which reflects the spread of civil disorder between geographic regions connected through social and communication networks. The results also expose the similarity between global social instability and the dynamics of natural hazards and epidemics.**

Civil unrest contagion occurs when social, economic, and political stress accumulate slowly, and is released spontaneously in the form of social unrest on short time scales to nearest and long-range neighboring regions that are susceptible to social, economic, and political stress (1-5). Unrest events have led to significant societal and cultural changes throughout history. Examples include the spread of discontent in France in 1848 that proliferated to most of Europe and parts of Latin America; the wave of urban racial riots in the United States in the 1960s; and the 1989 uprisings against communism in various central and eastern European countries, symbolized by the fall of the Berlin Wall. More recently, social instability has spread rapidly in the Arab world – from nonviolent protest movements in Tunisia and Egypt that toppled long-established authoritarian regimes, to a protest movement that evolved to a full-blown civil war in Libya. These social unrest events span the full spectrum from civil wars, revolutions, and coups d'état that have killed millions of people to relatively peaceful forms of intra-state conflicts, such as anti-government demonstrations, riots, and general strikes (5-10).

A pertinent question from large-scale social dynamics and policymaking standpoints is what causes the extent and outbreaks of civil unrest spreading. Social unrest has been attributed to a variety of social, political, economic, and environmental causes including racial and ethnic tensions (11), food scarcity and food price increases (6-10), variations in international commodity prices (12, 13), economic shocks (14), climate change and rainfall shocks (15, 16) and demographic changes (17). Despite these conditions, we show that external causes are not necessary to explain the observed magnitude of almost a century of riots and collective protests across the world. Instead, we provide a parsimonious explanation of social unrest dynamics based on a hypothesis that widespread unrest arises from internal processes of positive feedback and cascading effects in the form of contagion and social diffusion over spatially interdependent regions connected through social and mass communication networks. We analyze records of civil unrest events, compiled from newspaper reports, of 170 countries covering the period of 1919 through 2008 (see SI-1). The long-term event dataset analyzed here includes the number of incidents of three main indicators: anti-government demonstrations, riots, and general strikes. We group countries by their geographical region (see SI-2), and study the corresponding distributions of the number of civil unrest events per year. The results reported here (described later and shown in Fig. 3) also apply at the level of individual countries. Fig. 1 shows the size distributions of the number of unrest incidents (the sum over the three event indicators) for various geographical regions of the world (see SI-2). The data indicates that there is a wide variation in the characteristics of civil unrest with no apparent pattern of unrest dynamics in time or geographical space. Although the study of collective human dynamics, including collective aggression, has been the focus of much discussion in the context of modeling and identification of universal patterns of behavior (18-25), the mechanisms leading to this diverse behavior of social unrest are unclear, and have never been attempted before. Here, we consider social instability as a generalized spatial epidemics phenomenon, similar to other spatially extended dynamical systems in the physical and biological sciences, such as earthquakes, forest fires, and epidemics (26-28). Our model provides a parsimonious quantitative framework that is able to explain and reproduce the full range of empirical civil unrest event count distributions for all regions as shown in Fig. 1.

The model (shown schematically in Fig. 2) divides the country into sites ("urban clusters") placed on a two-dimensional grid. We assume that unrest activity is transmitted, with infectiousness probability $\mu$, along two kinds of links: short-range links between sites that are directly adjacent to each other in either the horizontal and vertical directions; and long-range links created with probability $m$ between each site and another site selected uniformly at random from the grid. The connectedness between sites on the grid reflects geographic proximity, social proximity, and proximity within the mass communication networks along which social instability is transmitted. Social, economic, and political stress accumulates slowly on the grid with probability $p$ per site ('unrest susceptibility' rate), which then become susceptible to unrest activity. Social unrest is released spontaneously with probability $f$ per susceptible site ('spontaneous outburst' rate), and is diffused on short time scales to nearest and distant susceptible sites. This activity can lead to further instabilities and avalanches of unrest events throughout the grid. We measure the simulated incidence of civil unrest by the number of sites that are involved in the spread of unrest activity.

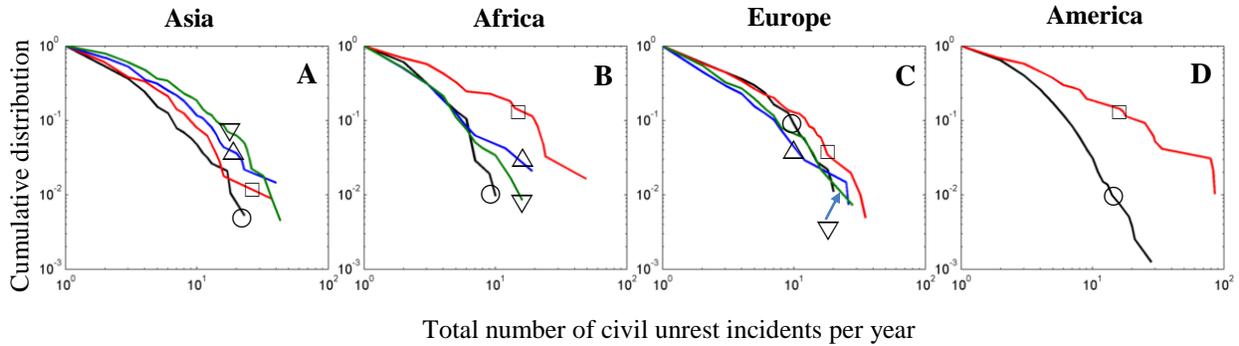

Total number of civil unrest incidents per year

**Figure 1| Observed civil unrest event count distributions.** We measure the incidence of civil unrest events per year by summing over the reported country-level number of anti-government demonstrations, riots, and general strikes (see SI-1), for all countries within a particular subregion of the world (see SI-2). The figure shows the log-log plot of the complementary cumulative distribution of civil unrest event count, $P(X \geq x)$. (**A**) Unrest event count distributions for geographical subregions of Asia: Western Asia (○); South-Eastern Asia (□); Eastern Asia (△); Southern Central Asia (▽). (**B**) Unrest event count distributions for geographical subregions of Africa: Western Africa (○); Southern Africa (□); Middle Africa (△); Eastern Africa (▽). (**C**) Unrest event count distributions for geographical subregions of Europe: Western Europe (○); Southern Europe (□); Northern Europe (△); Eastern Europe (▽). (**D**) Unrest event count distributions for geographical subregions of America: Caribbean, Central, and South America (○); North America (□).

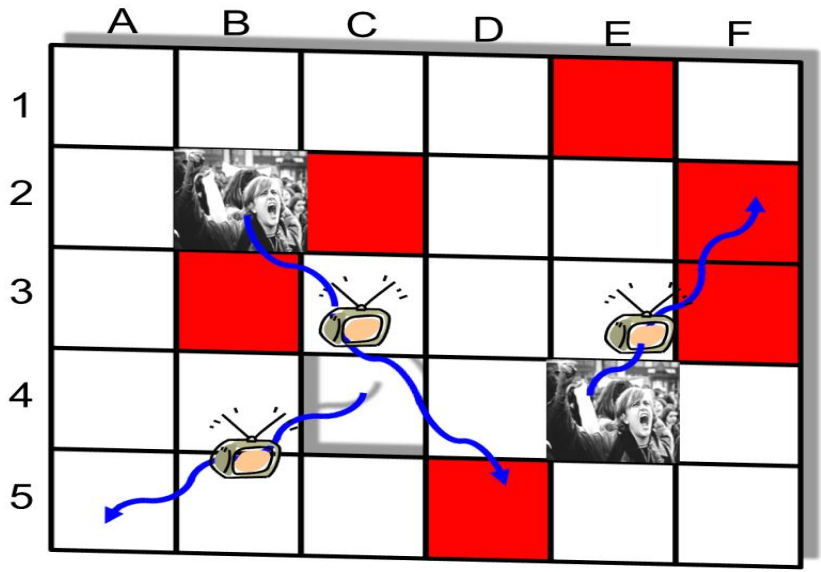

**Figure 2| Social unrest spatial contagion model.** The model is defined on a square grid of $N \times N$ sites, which represents the division of a country into urban clusters (see SI-4). Each site of

the grid can be in one of three different states: empty ("white"), susceptible to social unrest ("red"), and involved in social unrest ("crowd"). Sites are one grid step apart if they are directly adjacent to each other in either the horizontal or vertical direction, or are connected through weak links (e.g., site E4 is connected to site F2). The weak links are formed by associating with each site, with probability $m$, a single link to a site selected uniformly at random from the grid. The grid is updated synchronously according to the following rules: at each time step, empty sites become susceptible with probability $p$, and susceptible sites become involved in social unrest with probability $f$. Unrest contagion occurs on a short time scale as follows. If a site is involved in social unrest (e.g., sites B2 and E4), the unrest activity spreads with probability $\mu$ to susceptible sites that are one grid step apart (e.g., sites C2, B3, or D5), which in turn can lead (with probability $\mu$) to further instabilities of susceptible sites that are two grid steps apart, three grid steps apart, and so on. Each of the sites involved in social unrest during a time step contributes to the size of the unrest contagion.

We have specified a plausible size for the grid based on changes in the average population of a country over the period analyzed (1919 to 2008), and a characteristic urban cluster as defined by the U.S. Census Bureau (see SI-4). The outburst ($f$) and susceptibility ($p$) rates have been set such that $f \ll p \ll 1$, which reflects the time scale separation that often underlies riots, unrest and revolutions (5-10). This leaves only two free parameters: the probability $m$ of establishing long-range links between sites, and the infectiousness rate $\mu$ of transmitting social instability on short time scales to nearest and distant susceptible sites. Given a set of parameters, a computer simulation was run for a number of time steps (see SI-4), and the simulated distribution of the total unrest event count was determined. The free parameters associated with specific geographical regions of the world were chosen by minimizing the statistical distance between the simulated and empirical distributions (see SI-3). Using these parameters in the computer simulation model, we find that the model is able to reproduce the observed distributions remarkably well over the full range of the world's geographic regions (Fig. 3).

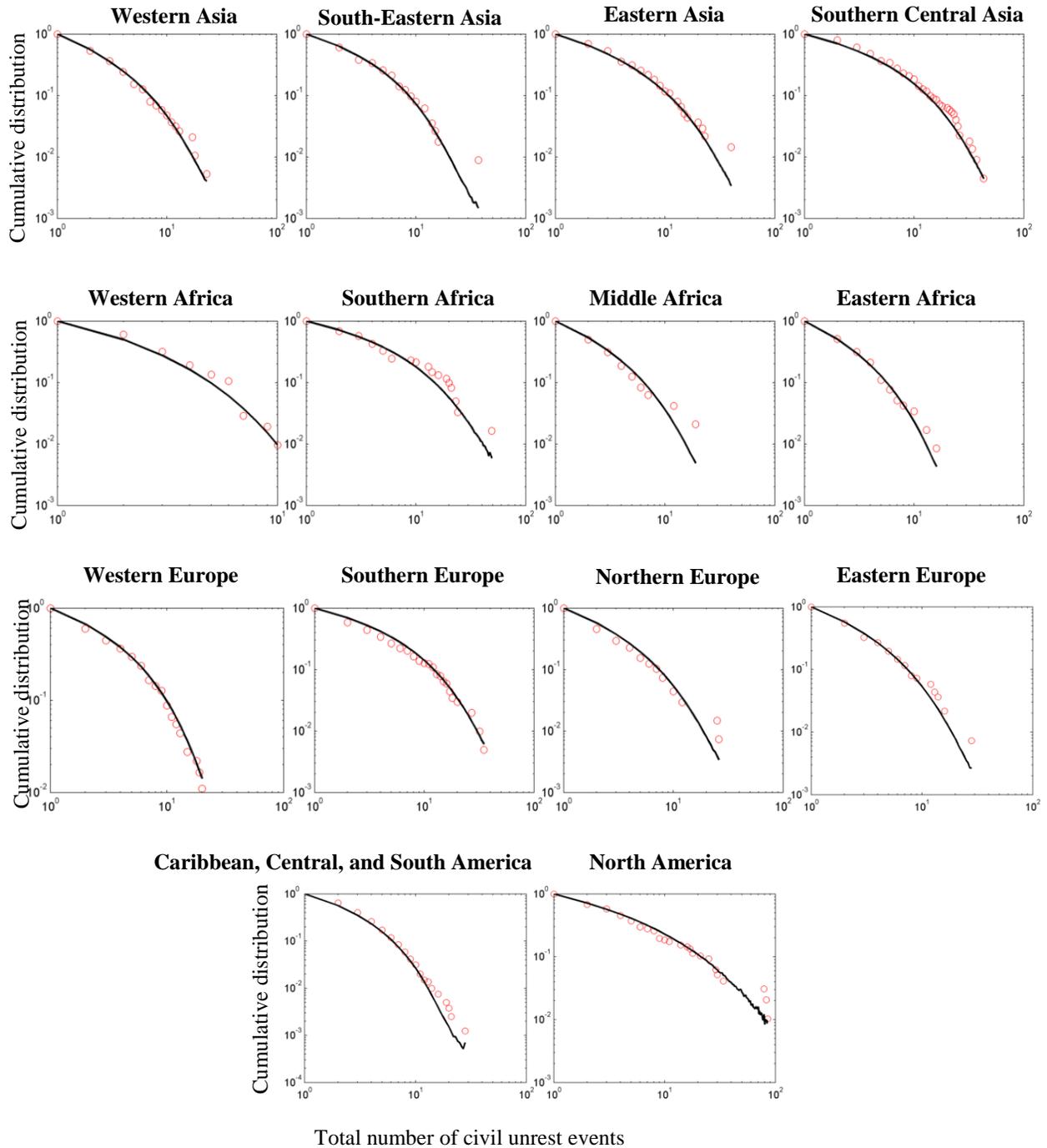

Total number of civil unrest events

**Figure 3| Observed data and best-fit curves for civil unrest event count distributions.** Observed values are denoted by circles. Solid best-fit lines denote average distributions calculated from 500 realizations of the social unrest contagion model. The goodness of fit of the model relative to the empirically observed unrest event count distributions was determined by measuring the distance between the observed and simulated distributions (SI-3). Here, we use the tail-weighted Kolmogorov–Smirnov (wKS) statistic (SI-4). The fit of the model is very good for all regions. Goodness-of-fit (wKS) and fitted parameters $(m, \mu)$ for all regions: Western Asia =

0.1269 ($m = 0.85$, $\mu = 0.16$), South-Eastern Asia = 0.1581 ($m = 0.2$, $\mu = 0.22$), Eastern Asia = 0.1492 ($m = 0.45$, $\mu = 0.22$), Southern Central Asia = 0.1078 ($m = 0.2$, $\mu = 0.26$), Western Africa = 0.1749 ($m = 0.1$, $\mu = 0.16$), Southern Africa = 0.2513 ($m = 0.2$, $\mu = 0.27$), Middle Africa = 0.1856 ($m = 0.9$, $\mu = 0.15$), Eastern Africa = 0.1191 ($m = 0.85$, $\mu = 0.14$), Western Europe = 0.1436 ($m = 0.05$, $\mu = 0.25$), Southern Europe = 0.11220 ($m = 0.15$, $\mu = 0.26$), Northern Europe = 0.16 ($m = 0.75$, $\mu = 0.17$), Eastern Europe = 0.1634 ($m = 0.75$, $\mu = 0.17$), Caribbean, Central, and South America = 0.0707 ($m = 0.1$, $\mu = 0.19$), North America = 0.2349 ($m = 0.5$, $\mu = 0.26$). Values of wKS that are less than 0.3 represent good fits (SI-3).

The unrest contagion model is sufficiently flexible to accommodate a wide range of possible unrest event count distributions. "Broad-scale" distributions that show a power–law regime with a sharp cutoff in the tail are obtained when the infectiousness rate $\mu = 1$, and when the outburst rate $f$ is very small relative to the susceptibility rate $p$ (i.e., $p/f \to \infty$), even in the absence of long-range connections ($m = 0$). ''Single-scale'' distributions with fast-decaying tails arise when the infectiousness rate $\mu \ll 1$, in which case augmenting the grid with a network of long-range connections ($m > 0$) could lead to broad-scale distributions. In general, different forms of social or communication networks that connect the regions of the country will generate different civil unrest event count distributions.

The unrest contagion patterns of each of the world's geographic regions (Fig. 1) are uniquely characterized by the parameters of the model (Fig. 3). Unlike critical phenomena where universality arises from the irrelevance of particular details of the system (29), here universality arises from the fact that social unrest contagion is governed by the same mechanisms despite idiosyncrasies of individual countries and geographic regions. The mechanisms we uncover separate the phenomenon of rioting and social instability into three time scales: the unrest infectiousness rate from disrupted regions to neighboring regions that are susceptible to social unrest, the rate by which regions become susceptible to unrest activity due to social, economic, and political stress, and the rate by which social unrest is released spontaneously in susceptible regions ($\mu \gg p \gg f$). The spatial contagion mechanism here arises from interdependence of closely related regions; people participate in collective protest because of long-standing social, economic, and political stress, and because others have recently done so. If rioters see others they might respond similarly even if their external conditions have not changed, and protests spreads across social networks and from place to place. While our parsimonious model does not prove that exogenous causes play no role in determining the intensity of civil unrest, it does say that exogenous causes are not necessary to explain the observed data, and that the pursuit of independent variables that predict the occurrence of civil unrest events in space and time may be illusory.

Our results have several practical implications. First, the parameters of the model can be estimated from unrest data that includes small and medium sized events, and then be used to quantify the risk of large-sized events. Second, monitoring the parameters of the model and trends in their values over time through comprehensive ongoing unrest data, may serve as an early warning signal for increased vulnerability to social instability.

Acknowledgements: I am grateful to Marcus A. M. de Aguiar, Zvi Bar-Yam, and Irving R. Epstein for helpful comments.

---